\documentclass[11pt]{article}
\usepackage{titlesec}
\usepackage[utf8]{inputenc}
\usepackage{color,setspace,graphicx,multirow}
\usepackage[top=2.4cm,left=2.4cm,top=2.4cm,bottom=2.4cm,includefoot]{geometry}
\usepackage{amsmath}
\usepackage{longtable}
\usepackage{authblk}
\usepackage[style=nature]{biblatex}
\usepackage{hyperref}
\bibliography{library}
\usepackage{titlesec}
\date{ }

\title{No complexity-stability relationship in natural communities}

\author[1]{Claire Jacquet}
\author[1,2]{Charlotte Moritz}
\author[2]{Lyne Morissette}
\author[1]{Pierre Legagneux}
\author[3]{Fran\c cois Massol}
\author[2]{Phillippe Archambault}

\author[1]{Dominique Gravel}

\affil[1]{Université du Québec à Rimouski, Département de biologie, chimie et géographie, 300 Allée des Ursulines, Québec, Canada G5L 3A1.}
\affil[2]{Université du Québec à Rimouski, Institut des sciences de la mer de Rimouski, 300 Allée des Ursulines, Québec, Canada. G5L 3A1.}
\affil[3]{Centre d'Ecologie Fonctionnelle et Evolutive, 1919 route de Mende, 34293 Montpellier cedex 5, France.}

\begin{document}

\maketitle

\begin{abstract}
Understanding the mechanisms responsible for the stability and persistence of natural communities is one of the greatest challenges in ecology \parencite{McCann2000}. Robert May showed that contrary to intuition, complex randomly built communities are less likely to be stable than simpler ones \parencite{May1972a, May2001}. For four decades, ecologists have tried to isolate the non-random characteristics of natural communities that could explain how they persist despite their complexity \parencite{McCann2000}. Surprisingly, few attempts have been tried to test May’s fundamental prediction and we still ignore if there is indeed a relationship between stability and complexity. Here, we performed a comparative stability analysis of 119 quantitative food webs sampled worldwide, from marine, freshwater and terrestrial habitats. Food webs were compiled using a standard methodology to build Ecopath mass-balance models. Our analysis reveals that classic descriptors of complexity (species richness, connectance and variance of interaction strengths) do not affect stability in natural food webs. Food web structure, which is far from random in real communities, reflects another form of complexity that we found influences dramatically the stability of real communities. We conclude that the occurrence of complex communities in nature is possible owing to their trophic structure.
\end{abstract}

The diversity-stability debate, initiated forty years ago \parencite{McCann2000}, stems from two apparently conflicting observations. On the one hand, complex communities are ubiquitous in nature, as illustrated by diverse tropical forests, coral reefs or intertidal communities, and it inspired ecologists to hypothesize that complexity could stabilize communities \parencite{MacArthur1955,Paine1966a}. On the other hand, a seminal mathematical analysis stated that complex systems are less likely to recover from small perturbations than simpler ones \parencite{May1972a}. This theoretical result was put forth by Robert May who studied the relationship between complexity and stability in random communities \parencite{May2001}. Community complexity was defined as the product of species diversity $S$, connectance $C$ and variance of interaction strengths $\sigma^2$. May predicted that a system could be stable only if the criterion $\sigma\sqrt{SC} <\bar{d}$ was satisfied, where $\bar{d}$ expresses the magnitude of intraspecific competition.

In an attempt to solve this paradox, a number of subsequent studies have shown that communities, and most notably food webs, have non-random structural properties and interaction strength distribution that promote their stability \parencite{Yodzis1981, Allesina2007a}. However, no clear consensus has emerged from this long debate and theoretical studies proposing alternative stabilizing network properties are continuously published (e.g. \parencite{Gravel2011a, Allesina2012}). The gap between theoretical and empirical investigations remains one of the main obstacles to resolve the debate. The first challenge is to get sufficiently good food web data with knowledge of trophic interactions and quantitative energy fluxes \parencite{Dunne2006b}. Another problem is that what theoreticians call “interaction strength” is generally not what empiricists measure in the field \parencite{Berlow2004}. Finally, the concept of stability itself encloses many different definitions \parencite{Grimm1997} and each of them yields a different diversity-stability relationship \parencite{Ives2007a}.

Here, we performed a local stability analysis on 119 quantitative food webs sampled worldwide from marine, freshwater and terrestrial habitats. The food webs were all compiled using a pre-defined methodology in order to use the Ecopath modeling framework \parencite{Christensen1992}. 
Ecopath is a trophic mass balance-model, the most widely used tool for ecosystem-based fisheries management, and has also been used to characterize unexploited ecosystems. A large amount of information is included in Ecopath models, such as trophic level, biomass, production and consumption rates of each species within a food web. Quantitative diet composition of species is also available, providing an accurate representation of trophic interactions within food webs. Ecopath models provide a unique opportunity to construct realistic community matrices with empirical data derived from a standardized protocol.

\begin{figure}[h!]
		\begin{center}
			\includegraphics[width=1\textwidth]{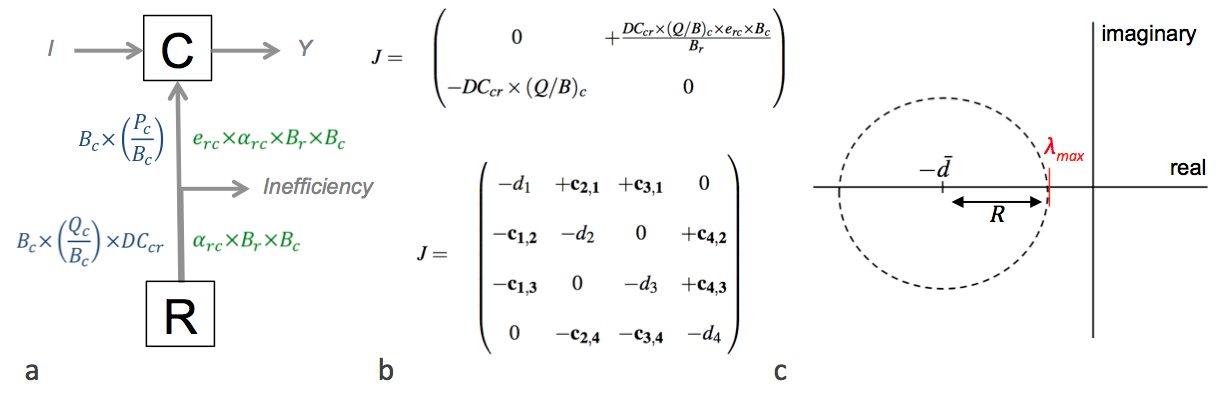}
		\end{center}
			\caption{\emph{Method summary: a) Equivalence of Ecopath and Lotka-Volterra models: simplified diagram of trophic flows between one consumer $C$ and one resource $R$ parameterized with Ecopath (in blue) and Lotka-Volterra (in green). $B$ is biomass $(t/km^2)$, $(P/B)_c$ and $(Q/B)_c$ are consumer production and consumption rates $/year$) respectively, $DC_{cr}$ is the proportion of resource $R$ in the diet of consumer $C$, $e_{rc}$ expresses the efficiency of a consumer to convert resource energy biomass with $e_{rc} = \frac{(P/B)_c}{(Q/B)_c}$, $I$ is the allochtonous input with $I = \omega \times (Q/B)_c \times B_c$, where $\omega$  is the proportion of input in the diet of the consumer, $Y$ represents the total consumer catches. b) Jacobian matrix construction: derivation of Jacobian matrix elements for the simplified food web presented in the diagram, and an example of Jacobian matrix structure observed in real food webs. c) Measure of stability: the eigenvalues of a Jacobian matrix are contained in a circle on the complex plane (axes cross at the origin). On the real axis, the maximum eigenvalue $Re(\lambda_{max})$ is influenced by the center of the circle $\bar{d}$, which is equal to the mean of intraspecific interaction terms (i.e. the diagonal elements of the Jacobian matrix), and by radius $R$, which is related to interspecific interaction terms (i.e. off diagonal elements of the Jacobian matrix) and is equal to $\sigma \sqrt{SC} $ in random communities.}}
	\end{figure}

The Ecopath model is canonical to the general Lotka-Volterra model that was studied by May \parencite{May1972a, May2001}, enabling us to study food web stability with traditional local stability analysis. Technically, the definition of local stability is that a system will come back to the equilibrium following a small perturbation. It does not guarantee stability following large perturbations (global stability), neither it quantifies persistence (the number of species remaining after a perturbation). We translated parameters of the Ecopath models into interaction coefficients of the Lotka-Volterra interaction model following the same approach as de Ruiter and colleagues \parencite{DeRuiter1995a} (Fig. 1a). The Lotka-Volterra model was slightly modified to account for the external exchanges of Ecopath models (see Appendix for further details).

Interaction coefficients from all pair-wise interactions of a food web constitute the interaction matrix \begin{math} {\bf A}=[\alpha_{ij}] \end{math}. Because of the Ecopath equilibrium assumption, a Jacobian matrix {\bf J} can be constructed for each food web by multiplying the interaction matrix {\bf A} with species biomass (Fig. 1b). We measured food web stability using the maximal eigenvalue (real part) of the Jacobian matrix in order to be directly comparable to May’s approach. This quantity indicates the rate with which a system returns to (if negative) or moves away from (if positive) equilibrium after small perturbations. Ecopath models rarely document the intraspecific interactions, we therefore centered for each web the distribution of eigenvalues on zero to avoid any bias in the evaluation of stability. Theory on Gerschgorin discs (a visual representation of the distribution of real and imaginary parts of eigenvalues for a given Jacobian matrix) states thats the mean real part is a function of the mean diagonal elements of the Jacobian matrix, while the range of eigenvalues is a function of off-diagonal elements \parencite{Haydon2000}. Adding intraspecific interactions (i.e. negative elements on the diagonal of the Jacobian matrix) has a stabilizing effect on food webs, by moving the mean of the eigenvalues toward negative values on the real axis (Fig. 1c). The centralization operation thus provides an unbiased estimation of the contribution of off-diagonal elements to the stability. Stability will decrease with increasing variance of the distribution of eigenvalues. 

We first investigated the relationship between stability and classic descriptors of community complexity  \parencite{Dunne2006b}, i.e. species richness $S$, connectance $C$, and variance of interaction strengths $\sigma^2$. This is the first time that the main prediction from May’s analysis is tested with empirical data. Contrary to theoretical prediction, we observed no relationship between food web stability and species diversity, neither with connectance nor with variance of interaction strengths (Fig. 2). Further analyses also reveal that this result is robust to the variability of sampling intensity among the 119 food webs (see Methods and Supplementary Information). The absence of a stability-complexity relationship we observed is  a striking departure from May's prediction, suggesting there are ecological processes or structural elements preventing the negative relationship between stability and complexity found in random communities.  We therefore investigated the mechanisms preventing this relationship to occur. 

	\begin{figure}[h!]
			\begin{center}
				\includegraphics[width=0.8\textwidth]{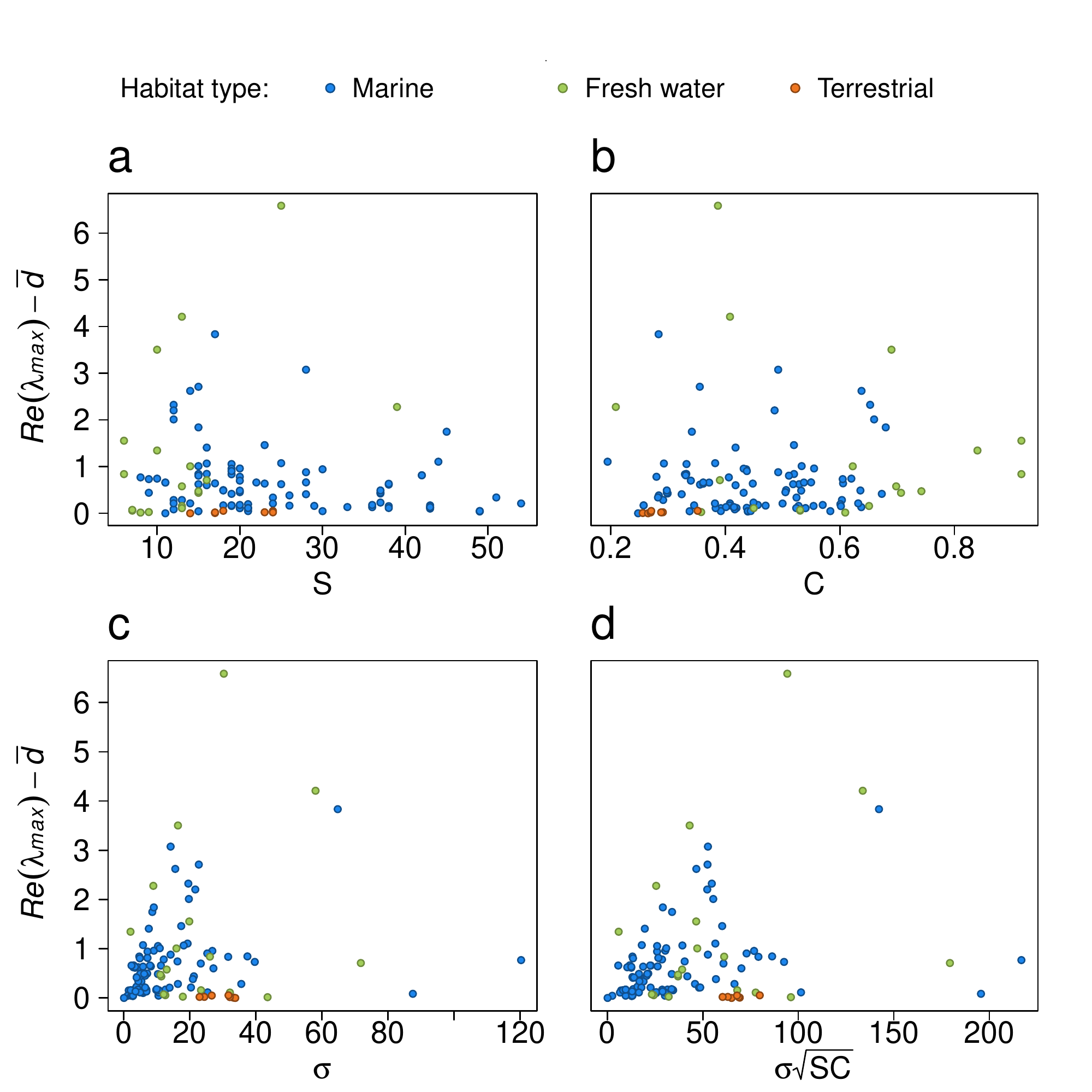}
			\end{center}
	\caption{\emph{ Food web stability related to measures of complexity in 119 food webs. a) Number of species $S (P = 0.10, R^2 < 0.01)$, b) Connectance $C = (L/S^2)$ with $L$ is the number of links $(P = 0.38, R^2 = 0.002)$, c) Standard deviation of interaction strengths $\sigma (P = 0.01, R^2 = 0.05)$, d) May’s complexity measure $\sigma \sqrt{SC}$ $(P = 0.002, R^2 = 0.07)$. Stability is measured as $Re(\lambda_{max}) - \bar{d}$ for marine (blue), freshwater (green) and terrestrial ecosystems (orange). Food webs with eigenvalues close to zero are the most stable. All quantities are dimensionless.}
	}
	\end{figure}

A consequence of the stability criterion $\sigma\sqrt{SC} <\bar{d}$ is that for complex systems to occur in nature, interaction strength should be weaker in species-rich, highly connected systems \parencite{May2001}. It is consistent with our observations: we found that the variance of interaction strengths  $\sigma^2$ across the 119 food webs was negatively correlated to the product of diversity and connectance  $\sqrt{SC}$ (Fig. 2). There is a growing constraint on interaction strength as diversity and connectance increase, which allows the overall complexity $\sigma\sqrt{SC}$ to remain relatively low and the system stable.

\begin{figure}[h!]	
			\begin{center}
				\includegraphics[width=0.5\textwidth]{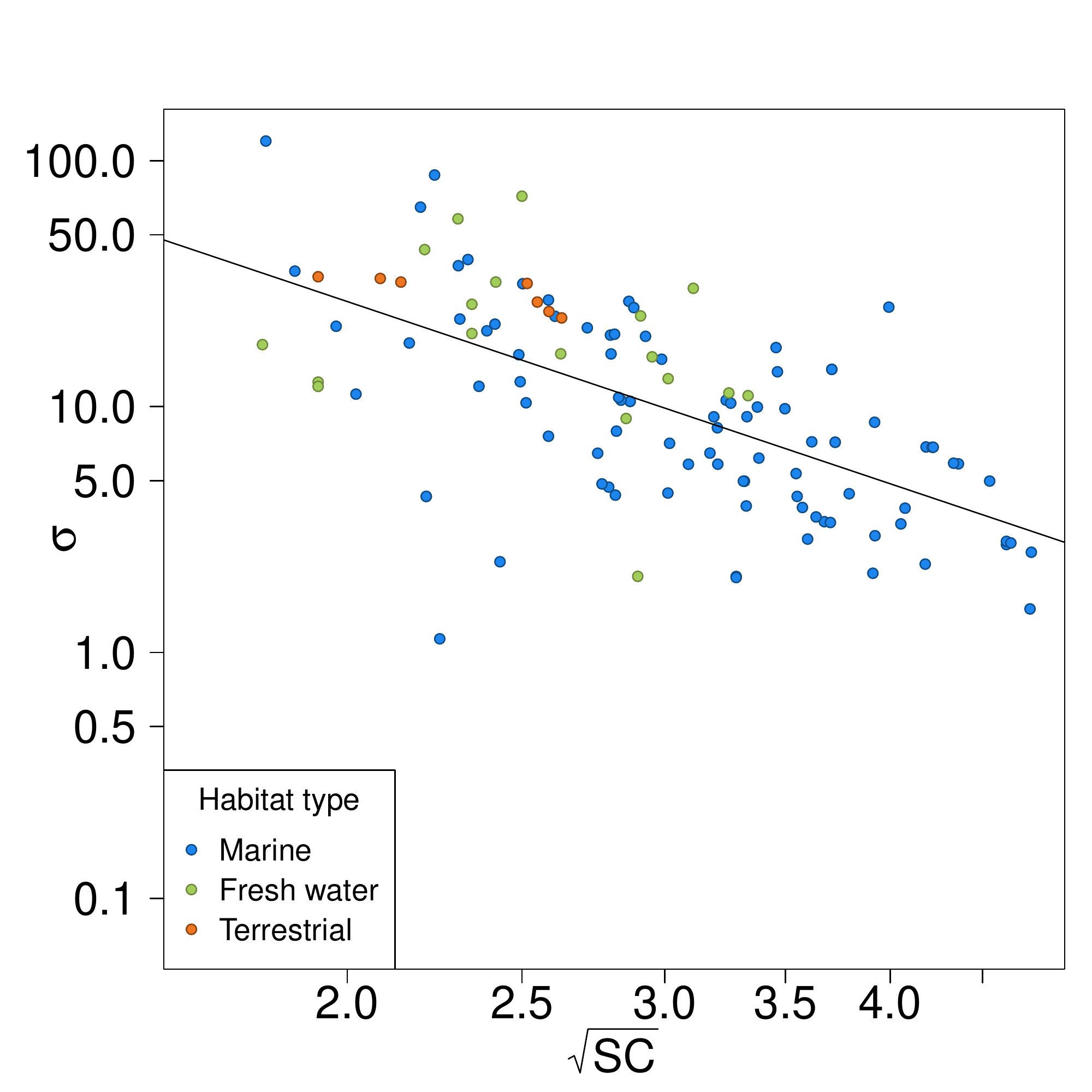}
			\end{center}
	\caption{ \emph{ Correlation between complexity parameters in real food webs, where $\sigma$ is the standard deviation of interaction strengths, $S$ the number of species and $C$ the connectance. The product $\sqrt{SC }$ were negatively correlated to $\sigma  (P<10^{-13}, R^2 = 0.39)$.}}
	\end{figure}

Theoretical studies of the stability-complexity relationship suggest that real communities have non-random structural properties promoting their stability despite their complexity \parencite{Dunne2006b}. We consequently looked at the existence of such properties and then investigated their respective contribution to stability by randomisations of some properties of the Jacobian matrices. \emph{H1: Interaction strength topology \parencite{Montoya2006a, DeAngelis1975}}. We  found that interaction strength was related to trophic level, the occurrence of strong interactions being more likely at low trophic levels. Moreover, there was a correlation between the magnitude of the Jacobian elements $c_{ij}$ and $c_{ji}$, since $c_{ji}= \frac{ - c_{ij} \times e_{ij} \times B_j }{B_i} $ (see Fig. 1b and Supplementary Information). We therefore hypothesized that food webs with random topology of interaction strengths (i.e. off-diagonal elements of the Jacobian matrix) are less stable than real food webs. \emph{H2: Interaction strength distribution \parencite{McCann1998, Berlow1999}}. In agreement with previous studies, we observed a leptokurtic distribution of interaction strengths (highly skewed towards many weak interactions). This pattern differs from May’s random community matrices in which interaction strengths were drawn from a normal distribution. Consequently, we hypothesized that food webs with a random frequency distribution of interaction strengths, illustrated by a normal distribution, are less stable than natural food webs. \emph{H3: Interaction type \parencite{May2001, Allesina2012, Stouffer2010a}}. Our dataset includes only predator-prey interactions and thus lacked other interaction types such as mutualism and interspecific competition. Based on previous findings, we assumed that communities with a strong proportion of mutualistic and competitive interactions are less stable than natural communities in which consumer-resource interactions prevail.

We performed randomisation tests to remove the structural properties of food web corresponding to our three hypotheses and computed stability of the permuted Jacobian matrices (see Methods for details). We employed this method to determine whether these structural properties had a significant effect on food web stability and to compare their respective stabilizing effect. The stability of the permuted food webs was compared to stability of the original food webs. We found that each of the three structural properties enhances food web stability (Fig.4). A remarkable feature however is their unequal contribution to stability. The type of interaction (with predator-prey module removal, H3) had the strongest impact on stability. Frequency distribution of interaction strengths, resulting in a large proportion of weak interactions (H2), was the second factor contributing to stability, followed by the topology of interaction strengths (H1 - Fig. 4).

	\begin{figure}[h!]		
			\begin{center}		
				\includegraphics[width=1\textwidth]{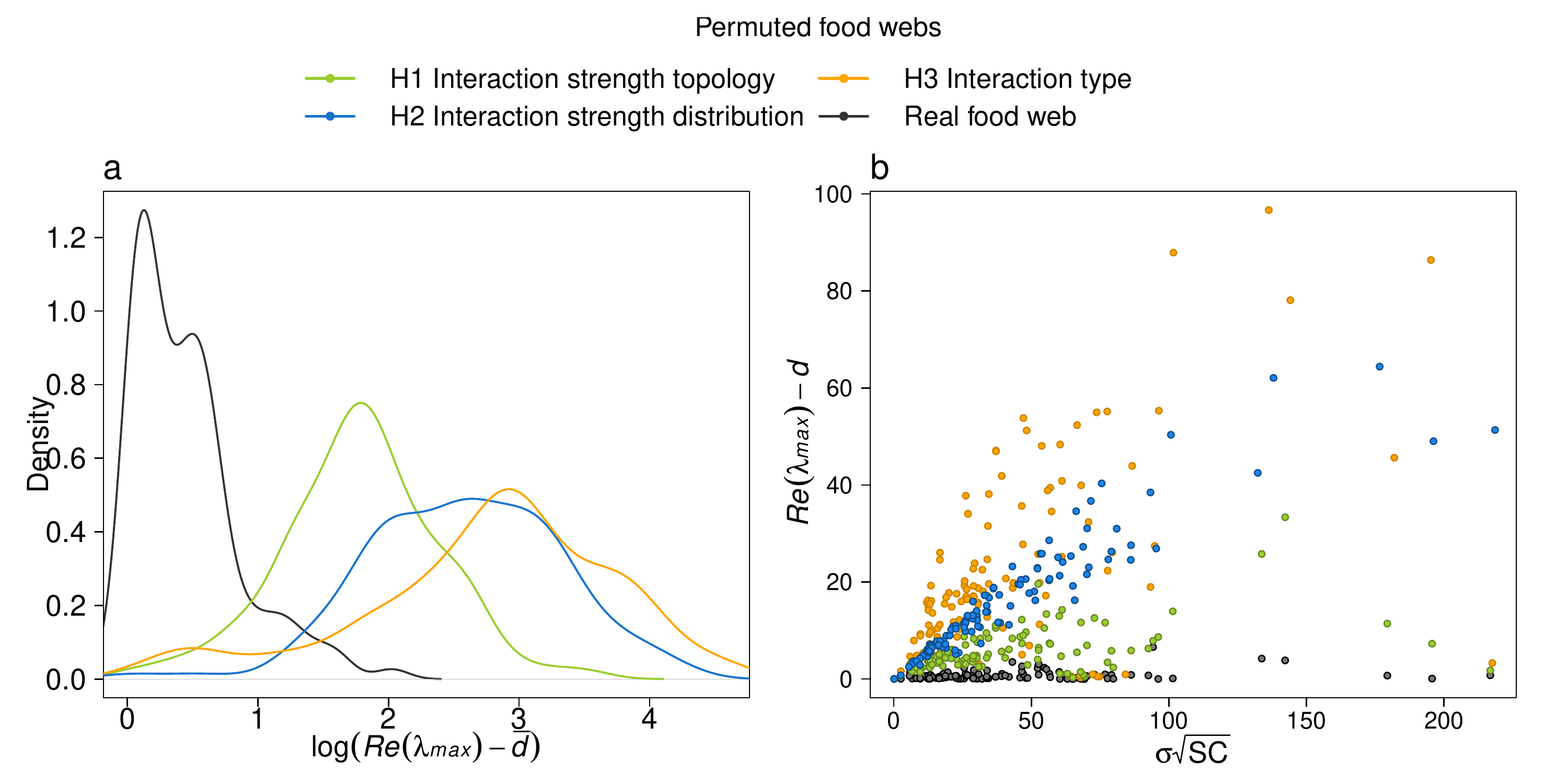}
			\end{center}		
	\caption{\emph{ a) Frequency distributions of eigenvalues of real and permuted food webs. Eigenvalues are on a logarithmic scale and dimensionless. Permutation tests were carried out 1000 times for each food web. Eigenvalue distributions were smoothed using a kernel density estimate of 0.28. b) Stability of real and permuted food webs related to May's complexity criterion. Real food webs $(P = 0.002, R^2 < 0.07)$, H1: interaction strength topology $(P<10^{-5}, R^2 = 0.17)$, H2: interaction strength distribution $(P<10^{-15}, R^2 = 0.86)$ and H3: interaction type $(P = 0.88, R^2 = 0.3)$. Stability is measured as $Re(\lambda_{max}) - \bar{d}$ and $\sigma \sqrt{SC} $ corresponds to May's stability criterion.  Communities with eigenvalues close to zero are the most stable}}	
	\end{figure}
	
Even if we conserved the same $S$, $C$ and $\sigma$, we found a negative complexity-stability relationship for all randomized Jacobian matrices (Fig. 5). The relationship was strongest when we normalised the interaction strength distribution (H2). We conclude that May's stability criterion do not apply to natural communities because of their non-random trophic structure, which has several stabilizing properties. First, the asymmetric sign structure (predator-prey modules) confers stability to Jacobian matrices (H3), as we found that for similar variation of interaction strengths, mutualistic and competitive interactions were destabilizing. Second, the frequency distribution of interaction strengths balanced the destabilizing effect of species richness. Interestingly, we observed in real data a strong negative correlation between the kurtosis $\kappa$ (index of the flatness/peakness of the interaction strength distribution) and species richness in real food webs. Thus the probability of having many weak interactions increases with species richness (Fig 3 and Supplementary Information). Finally, the nonrandom topology of interaction strengths along trophic level was also stabilizing, as suggested by previous studies \parencite{DeRuiter1995a, Neutel2002}.

The relevance of local stability analysis to study real communities may be questioned. More general and realistic definitions of stability have been introduced during the “complexity-stability” debate, such as persistence, resilience or resistance \parencite{Grimm1997}. Indeed, local stability analysis only tests the impact of small perturbations on ecological dynamics, and may not apply to large perturbations typical of most empirical studies. However, it allows the use of analytically tractable Jacobian matrices, and thus the investigation of May’s complexity-stability relationship on real communities.

Although May’s conclusion that diversity begets instability in random communities was correct, we showed that diversity is not related to stability in natural communities, a question that has stimulated ecological research for four decades. We found that intrinsic energetic organization of food webs is highly stabilizing and allow complex communities to recover from perturbations. The structural complexity of food webs occurs from the successive addition of consumers having an increasingly large diet, which causes a growing frequency of weak interactions.
The diversity-stability debate has contributed to the development of productive research that have pointed out the key role of the structural properties of real communities. There is increasing evidence that the strengths of trophic interactions are related to the body size distribution of species \parencite{Emmerson2004a}. Changes in frequency distributions of body size could therefore strongly affect community stability and resilience.

\printbibliography{}

\section{Methods}

{\bf Calibration of Ecopath models.} Ecopath provides a quantitative overview of how species interact in a food web. Species sharing the same prey and predators and having similar physiological characteristics are aggregated in trophic species. Ecopath rely on a system of linear equations decribing the in and out flows of each compartment. We compiled 119 Ecopath food web models from published studies. A list of models with references, habitat types, species richness $S$, connectance $C$ and standard deviation of interaction strengths $\sigma$ is available in Supplementary Information. Model calibration is based on the following input data: biomass, production rates, consumption rates, fishery yields, and diet composition for species of the food web. Input data can have different origins: field sampling (e.g. trawl survey), derived from similar Ecopath models, or known empirical relationships.

{\bf Parameterization of Lotka Volterra interaction coefficients.} We used the method from ref \parencite{DeRuiter1995a} to derive the Jacobian matrices from Ecopath models. Assuming direct dependence of feeding rates on predator population density, we calculated the per capita effect of predator $j$ on the growth rate of prey $i$ as $\alpha_{ij} = - \frac { (Q/B)_j \times DC_{ji} } {B_i}$ where $B$ is biomass $(t/km^2)$, $(P/B)_j$ and $(Q/B)_j$ are predator production and consumption rates ($/year$) respectively, $DC_{ji}$ is the proportion of species $i$ in the diet of predator $j$. Effects of prey on their predator are defined as predator growth resulting from this predation. Consequently, effect of the prey $i$ on the predator $j$ is related to effect of the predator on the prey according to: $\alpha_{ji} = - e_{ij} \times \alpha_{ij} $, where $e_{ij}$ is the efficiency with which $j$ converts food into biomass, from feeding on $i$: $e_{ij} = \frac{(P/B)_i}{(Q/B)_j}$.

{\bf Evaluation of Ecopath model quality.} We assessed the robustness of our comparative analysis to ensure that results were not an artifact of differences in model quality. The amount of aggregation of each model was measured, based on the criterion that groups with taxonomic name were more resolved than groups with trophic function names. We defined four resolution levels and qualified it for each trophic species with the following index values: taxonomic species (i.e. greenland turbot, index = 1), family/class (i.e. whales, gadoids; index = 0.7), trophic function (i.e. small demersal fish; index = 0.4) and general name (i.e. benthos, fish; index=0.1). Resolution indices $RI$ of Ecopath models correspond to the mean resolution index of species within each food web. We investigated the complexity-stability relationship on a subset of the 37 best resolved models with $RI \geq 0.7$. Results were similar to the overall analysis. Resolution indices and results of the stability analysis on the most resolved models are available in Supplementary Information.

{\bf Randomisation tests.} Reported maximal eigenvalues of randomised food webs correspond to the mean of 1000 replicates. For randomisation of interaction strength topology (H1), we permuted off-diagonal elements of the Jacobian matrix, keeping sign structure and the frequency distribution of interaction strengths, positive and negative terms were permuted separately in order to keep the initial interaction strength mean. For randomisation of interaction strength distribution (H2), we created a random Jacobian matrix in which off-diagonal elements were picked from a normal distribution $N (\mu, \sigma^2$) where $\mu$ is the mean and $\sigma^2$ the variance of initial Jacobian matrix elements, then we imposed the sign structure of the initial Jacobian matrix. Positive and negative terms of the initial Jacobian matrix were replaced by positive and negative normalised terms respectively. For the randomisation of sign structure (H3), we permuted only non-zero elements of the Jacobian matrix in order to keep the same link density and connectance than the initial Jacobian matrix, and a similar frequency distribution of the elements of the Jacobian matrix.

\end{document}


\maketitle
\tableofcontents{}

\clearpage

\section{Supplementary figures}

\subsection{Structural properties observed in food webs and corresponding null hypotheses (red lines).}
	\begin{figure}[!h]
		\centering\includegraphics[width=0.57\textwidth]{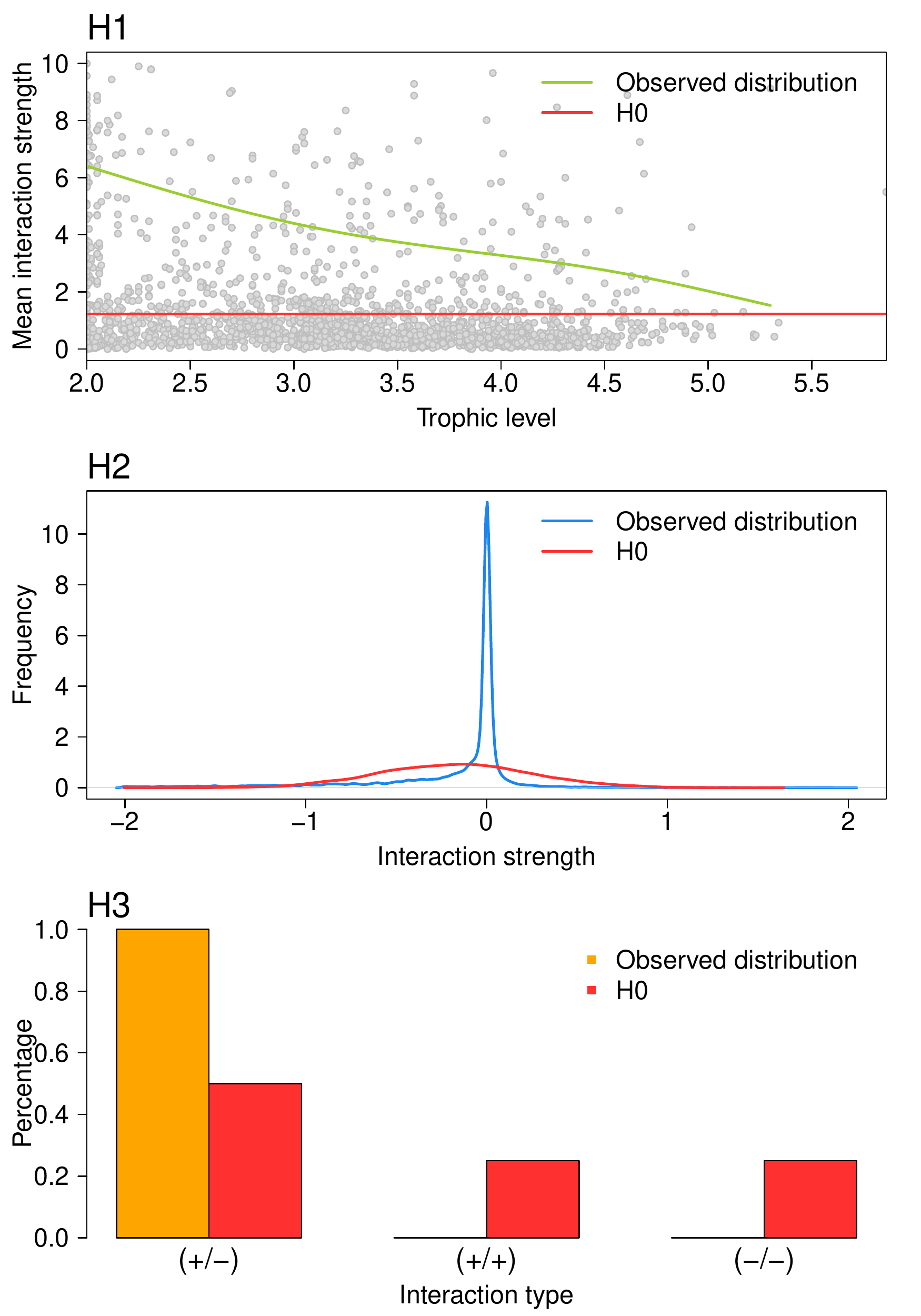}
	\caption{{\bf H1)} Species average interaction strength as a function of trophic level (top 5\% were omitted from the figure for illustration). Green line 95\% quantile regression illustrates average interaction strength pyramidal pattern. {\bf H2)} Frequency distribution of interaction strengths (only the interval [-2; 2] is represented, which corresponds to 90\% of all interaction coefficients). {\bf H3)} Interaction type frequency in real food web (orange) and in random food web (red) where trophic interactions $(+/-)$ or $(-/+)$, mutualism $(+/+)$ and interspecific competition $(-/-)$ have the same probability of occurrence.}
	\end{figure}

\subsection{Food web stability related to measures of complexity in the 37 best resolved models}
We were concerned that the lack of a complexity-stability relationship may have arisen because of the use of trophic species in food web models, inducing a bias in the evaluation of $S$ and consequently $C$ and $\sigma$. Indeed, species aggregation in trophic groups was unequal between Ecopath models, possibly hiding a relationship between diversity and stability,. We thus performed a stability analysis on a subset of the data, selecting only the most highly resolved models with low aggregation. Results were consistent with those of the overall analysis, demonstrating the robustness of the method to species aggregation. Aggregation level of Ecopath models was measured, based on the criterion that groups with taxonomic names were more resolved than groups with trophic function names.

\begin{figure}[!h]
	\centering\includegraphics[width=0.8\textwidth]{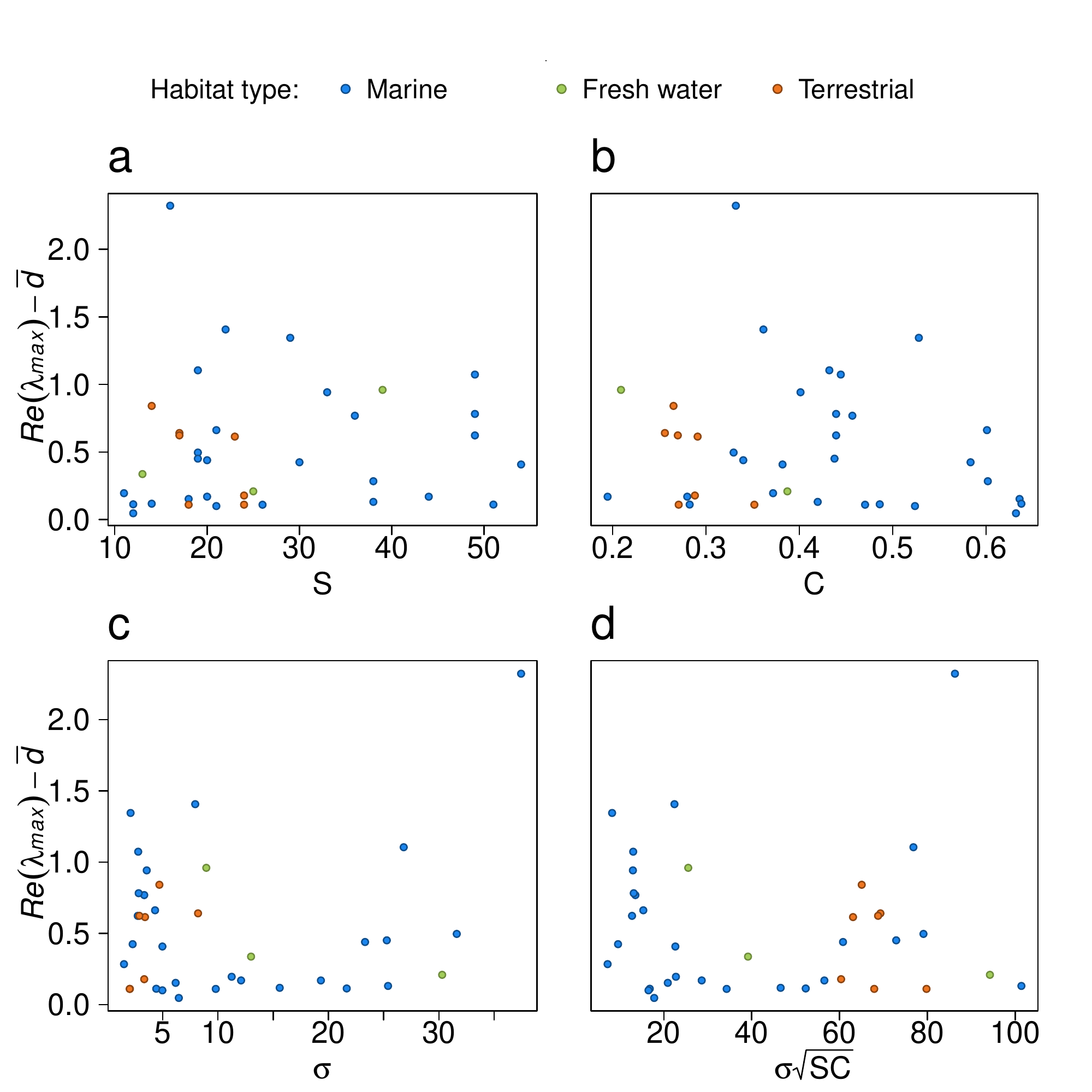}
	\caption{{\bf a)} Number of species $S (P=0.41, R^2=0.008)$, {\bf b)}  Connectance $C = \frac{L} {S^2}$ where $L$ is the number of links $(P=0.92, R^2=0.03)$, {\bf c)}  Standard deviation of interaction strengths $\sigma (P=0.16, R^2=0.03)$, {\bf d)}  May’s complexity measure $\sigma \sqrt{SC} (P=0.05, R^2=0.02)$. Stability is measured as the maximum eigenvalue of each food webs, for marine (blue), freshwater (green) and terrestrial ecosystems (orange). Food webs with eigenvalues close to zero are the most stable. All quantities are dimensionless.}
\end{figure}

\subsection{Correlation between species richness and the shape of interaction strength distribution}

\begin{figure}[h]
	\centering\includegraphics[width=0.8\textwidth]{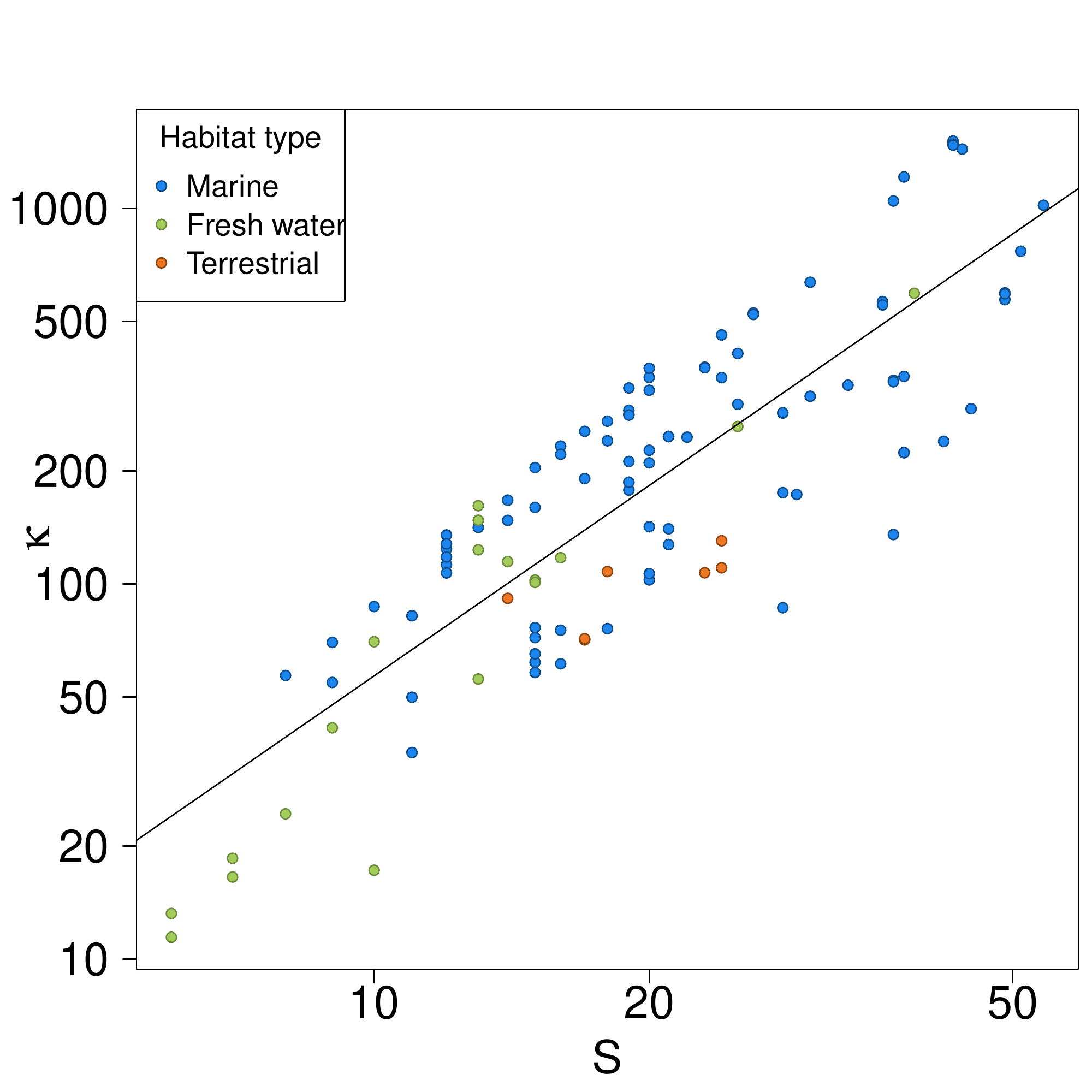}
		\caption{The kurtosis $\kappa$ is a measure of the flatness/peakness of the interaction strength distribution, for a normal distribution $\kappa = 0$. Species richness $S$ were positively correlated to $\kappa$ $(P<10^{-15}, R^2=0.71)$ on a logarithmic scale. All quantities are dimensionless.}

\end{figure}

\newpage
\section{Ecopath models}

\subsection{Dataset information}
Table S1: List of the 119 Ecopath models with references, habitat types, number of species $S$, connectance $C$, standard deviation of interaction strengths $\sigma$, and resolution index $RI$.
\begin{longtable}[c]{p{4.5cm} p{5cm} p{1.5cm} p{0.5cm} p{1cm} p{1cm} p{1cm}}

\hline
\textbf{Model Name} & \textbf{Reference}&\textbf{Habitat} & \textbf{S} & \textbf{C} & \textbf{$\sigma$} & \textbf{RI} \\
    \hline
      \endfirsthead 
\multicolumn{7}{c}%
{ \tablename\ \thetable{} -- continued from previous page} \\
\hline
\textbf{Model Name} & \textbf{Reference}&\textbf{Habitat} & \textbf{S} & \textbf{C} & \textbf{$\sigma$} & \textbf{RI} 
      \endhead 
    Alaka Prince William Sound OM & Okey \& Pauly 1999 \parencite{Okey1999} & marine & 18    & 0.35  & 12.63 & 0.62 \\
    Alto Golfo De California & Morales-Zárate et al. 2004 \parencite{Morales-Zarate2004}& marine & 28    & 0.55  & 2.99  & 0.63 \\
    Antarctica Weddel Sea & Jarre-Teichmann et al. 1997 \parencite{Jarre-Teichmann1997} & marine & 19    & 0.26  & 4.31  & 0.57 \\
    Arctic islands, Alert & Legagneuxet et al. (submitted)  & terrestrial & 17    & 0.26  & 33.22 & 0.86 \\
    Arctic islands, Bylot & Legagneux et al. (submitted)& terrestrial & 18    & 0.35  & 31.71 & 0.85 \\
    Arctic islands, Erkuta & Legagneux et al. (submitted)& terrestrial & 23    & 0.29  & 24.38 & 0.84 \\
    Arctic islands, Herschel & Legagneux et al. (submitted)& terrestrial & 24    & 0.27  & 26.63 & 0.85 \\
    Arctic islands, Nenetsky & Legagneux et al. (submitted)& terrestrial & 24    & 0.29  & 22.95 & 0.83 \\
    Arctic islands, Svalbard & Legagneux et al. (submitted)& terrestrial & 14    & 0.27  & 33.76 & 0.76 \\
    Arctic islands, Zackenberg & Legagneux et al. (submitted)& terrestrial & 17    & 0.27  & 32.11 & 0.86 \\
    Bali Strait & Buchary et al. 2002 \parencite{Buchary2002}& marine & 13    & 0.6   & 16.38 & 0.51 \\
    Bay Of Biscay 1998 & Ainsworth et al. 2001 \parencite{Ainsworth2001}& marine & 36    & 0.64  & 2.56  & 0.42 \\
    Bay Of Somme & Rybarczyk et al. 2003 \parencite{Rybarczyk2003} & marine & 8     & 0.41  & 120.28 & 0.33 \\
    Black Sea & Orek 2000 \parencite{Orek2000} & marine & 9     & 0.43  & 21.23 & 0.43 \\
    Brunei Darussalam, China Sea & Silvestre et al. 1993 \parencite{Silvestre1993}& marine & 12    & 0.65  & 19.54 & 0.45 \\
    Campeche Bank, Golf of Mexico & Vega-Cendejas et al. 1993 \parencite{Vega-Cendejas1993} & marine & 18    & 0.51  & 7.09  & 0.6 \\
    Campeche Sound & Manickchand-Heileman et al. 1998 \parencite{ManickchandHeileman1998}& marine & 24    & 0.52  & 5.34  & 0.61 \\
    Cape Verde & Stobberup et al. 2004 \parencite{Stobberup2004}& marine & 24    & 0.5   & 13.86 & 0.61 \\
    Caribbean & Morissette et al. 2010 \parencite{Morissette2010}& marine & 28    & 0.32  & 4.45  & 0.68 \\
    Celestun & Chavez et al. 1993 \parencite{Chavez1993} & marine & 15    & 0.53  & 4.37  & 0.6 \\
    Central Atlantic 50s & Vasconcellos \& Watson 2004 \parencite{Vasconcellos2004} & marine & 37    & 0.3   & 3.95  & 0.56 \\
    Central Chile 1992 & Neira et al. 2004 \parencite{Neira2004} & marine & 20    & 0.28  & 12.1  & 0.93 \\
    Central Gulf Of California & Arreguin-Sanchez et al. 2002 \parencite{Arreguin-Sanchez2002} & marine & 25    & 0.38  & 5.83  & 0.63 \\
    Central Pacific, sharks & Kitchell et al. 2002 \parencite{Kitchell2002}  & marine & 21    & 0.6   & 4.31  & 0.8 \\
    Chesapeake Present & Christensen et al. 2009 \parencite{Christensen2009} & marine & 44    & 0.19  & 19.32 & 0.79 \\
    Darwin Harbour, Australia & Martin 2005 \parencite{Martin2005} & marine & 20    & 0.51  & 6.47  & 0.5 \\
    Eastern Scotian Shelf 90s & Bundy 2004 \parencite{Bundy2004} & marine & 38    & 0.6   & 1.5   & 0.72 \\
    Eastern Tropical Pacific & Olson \& Walters 2003 \parencite{Olson2003} & marine & 38    & 0.42  & 25.39 & 0.81 \\
    Etang de Thau, France & Palomares et al. 1993 \parencite{Palomares1993} & freshwater & 10    & 0.84  & 2.04  & 0.52 \\
    Gambia 1986 & Mendy 2003 \parencite{Mendy2003} & marine & 21    & 0.39  & 10.5  & 0.63 \\
    Gambia 1992 & Mendy 2004 \parencite{Mendy2004} & marine & 21    & 0.39  & 10.5  & 0.63 \\
    Gambia 1995 & Mendy 2004 \parencite{Mendy2004} & marine & 21    & 0.38  & 10.63 & 0.63 \\
    Gironde Estuary, France & Lobry 2004 \parencite{Lobry2004} & marine & 16    & 0.42  & 7.58  & 0.56 \\
    Golfo Dulce, Costa Rica & Wolff et al. 1996 \parencite{Wolff1996} & marine & 20    & 0.56  & 9.1   & 0.6 \\
    Guinee 1985 & Guénette \& Diallo 2004 \parencite{Guenette2004}  & marine & 43    & 0.44  & 5.85  & 0.66 \\
    Guinee 1998 & Guénette \& Diallo 2004 \parencite{Guenette2004}  & marine & 43    & 0.44  & 5.89  & 0.66 \\
    Guinee Bissau 1991 & Amorim et al. 2003 \parencite{Amorim2003} & marine & 30    & 0.44  & 7.19  & 0.57 \\
    Gulf of Salamanca, Upwelling & Duarte \& Garcia 2004 \parencite{Duarte2004} & marine & 17    & 0.61  & 8.2   & 0.54 \\
    High Barents Sea AllJuvs1990 & Blanchard et al. 2002 \parencite{Blanchard2002} & marine & 15    & 0.52  & 4.71  & 0.54 \\
    High Barents Sea Final 1990 & Blanchard et al. 2002  \parencite{Blanchard2002} & marine & 38    & 0.36  & 3.4   & 0.66 \\
    Huizache Caimanero, Mexico & Zetina-Rejón et al. 2004  \parencite{Zetina-Rejon2004} & marine & 25    & 0.52  & 2.89  & 0.64 \\
    Iceland Fisheries & Buchary 2001  \parencite{Buchary2001} & marine & 20    & 0.54  & 2.04  & 0.69 \\
    Icelandic Shelf & Mendy \& Buchary 2001  \parencite{Buchary2001} & marine & 20    & 0.54  & 2.02  & 0.69 \\
    Jalisco y Colima & Galván-Piña 2005  \parencite{V.H.2005} & marine & 36    & 0.46  & 3.33  & 0.7 \\
    Kuala Terengganu, Malaysia & Liew \& Chan 1987  \parencite{Liew1987} & marine & 12    & 0.66  & 19.71 & 0.4 \\
    Kuosheng Bay, Taiwan & Lin et al. 2004  \parencite{Lin2004} & marine & 16    & 0.29  & 18.15 & 0.49 \\
    Lagoon Chiku, Taiwan & Lin et al. 1999  \parencite{Lin1999} & marine & 12    & 0.42  & 87.51 & 0.5 \\
    Lagoon of Venice & Carrer \& Opitz 1999  \parencite{Opitz1996} & marine & 15    & 0.34  & 1.14  & 0.48 \\
    Laguna De Bay, Philippines, 1950 & Delos Reyes 1995  \parencite{DelosReyes1995} & marine & 20    & 0.34  & 23.31 & 0.73 \\
    Laguna De Bay, Philippines, 1980 & Delos Reyes \& Martens 1993  \parencite{DelosReyes1994} & marine & 16    & 0.33  & 37.43 & 0.76 \\
    Laguna De Bay, Philippines, 1990 & Delos Reyes 1995 \parencite{DelosReyes1995}  & marine & 19    & 0.33  & 31.61 & 0.73 \\
    Lake Aydat, France & Reyes-Marchant et al. 1993 \parencite{ReyesMarchant1993}  & freshwater & 10    & 0.69  & 16.4  & 0.49 \\
    Lake Chad, Africa & Palomares et al. 1993 \parencite{Palomares1993} & freshwater & 14    & 0.62  & 15.94 & 0.55 \\
    Lake George, Uganda & Moreau et al. 1993  \parencite{Moreau1993} & freshwater & 13    & 0.7   & 13    & 0.72 \\
    Lake Kariba, Africa & Machena et al. 1993  \parencite{Machena1993} & freshwater & 9     & 0.36  & 17.88 & 0.67 \\
    Lake Kinneret, Israel & Walline et al. 1993  \parencite{Walline1993} & freshwater & 13    & 0.41  & 58.05 & 0.52 \\
    Lake Malawi 2, Africa & Degnbol 1993  \parencite{Degnbol1993} & freshwater & 8     & 0.61  & 43.53 & 0.63 \\
    Lake Malawi, Africa & Nsiku 1999  \parencite{Nsiku1999} & freshwater & 25    & 0.39  & 30.29 & 0.7 \\
    Lake Tanganyka, Africa, 1975 & Moreau et al. 1993a  \parencite{Moreau1993a} & freshwater & 6     & 0.92  & 26.08 & 0.5 \\
    Lake Tanganyka, Africa, 1981 & Moreau et al. 1993a  \parencite{Moreau1993a} & freshwater & 6     & 0.92  & 19.83 & 0.5 \\
    Lake Turkana, Kenya, 1973 & Kolding 1993  \parencite{Kolding1993} & freshwater & 7     & 0.53  & 12.59 & 0.57 \\
    LakeTurkana, Kenya, 1987 & Kolding 1993  \parencite{Kolding1993} & freshwater & 7     & 0.53  & 12.09 & 0.57 \\  
    Lake Victoria, Africa, 1971 & Moreau et al. 1993b  \parencite{Moreau1993b} & freshwater & 15    & 0.71  & 11.37 & 0.62 \\
    Lake Victoria, Africa, 1985 & Moreau et al. 1993b  \parencite{Moreau1993b} & freshwater & 15    & 0.74  & 11.09 & 0.62 \\
    Looe Key, Florida, USA & Venier \& Pauly 1997  \parencite{Venier1997} & marine & 19    & 0.55  & 10.62 & 0.51 \\
    Low Barents Sea 1995 & Blanchard et al. 2002  \parencite{Blanchard2002} & marine & 38    & 0.36  & 3.38  & 0.69 \\
    Low Barents Sea Juvs 1995 & Blanchard et al. 2002  \parencite{Blanchard2002} & marine & 15    & 0.51  & 4.85  & 0.54 \\
    Mandinga Lagoon, Mexico & De la Cruz-Aguero 1993  \parencite{Cruz-Aguero1993} & marine & 19    & 0.33  & 10.37 & 0.65 \\
    Maputo Bay, Mozambique & De Paula et al. 1993  \parencite{Silva1993} & marine & 9     & 0.6   & 39.66 & 0.5 \\
    Mid Atlantic Bight & Okey 2001 \parencite{Okey2001}& marine & 54    & 0.38  & 4.98  & 0.7 \\
    Monterey Bay, California & Olivieri et al. 1993 \parencite{Olivieri1993} & marine & 15    & 0.36  & 22.71 & 0.36 \\
    Moorea Barrier reef & Arias- González et al. 1997 \parencite{AriasGonzales1997} & marine & 45    & 0.34  & 8.64  & 0.54 \\
    Moorea Fringing reef & Arias- González et al. 1997 \parencite{AriasGonzales1997} & marine & 42    & 0.33  & 7.17  & 0.56 \\
    Morocco 1984 & Stanford et al. 2001 & marine \parencite{Stanford2004} & 37    & 0.45  & 3.87  & 0.52 \\
    Newfoundland Grand Banks 1900 & Heymans \& Pitcher 2002b \parencite{Heymans2002b} & marine & 49    & 0.44  & 2.75  & 0.72 \\
    Newfoundland Grand Banks mid-1980s & Bundy 2001 \parencite{Bundy2001} & marine & 30    & 0.58  & 2.29  & 0.76 \\
    Newfoundland Grand Banks mid-1980s & Heymans \& Pitcher 2002a \parencite{Heymans2002a} & marine & 49    & 0.44  & 2.83  & 0.76 \\
    Newfounland  Grand Banks mid-1990s & Heymans \& Pitcher 2002a \parencite{Heymans2002a} & marine & 49    & 0.44  & 3.87  & 0.76 \\
    North Atlantic 1950s & Vasconcellos \& Watson 2004 \parencite{Vasconcellos2004}  & marine & 37    & 0.3   & 4.97  & 0.59 \\
    North Atlantic 1990s & Vasconcellos \& Watson 2004 \parencite{Vasconcellos2004} & marine & 37    & 0.3   & 4.98  & 0.59 \\
    Northeastern Venezuela shelf & Mendoza 1993 \parencite{Mendoza1993} & marine & 15 & 0.57 & 23.47 & 0.54 \\
    Northwest Africa & Morissette et al. 2010a \parencite{Morissette2010a} & marine & 26    & 0.28  & 20.92 & 0.69 \\
    Orbetello Lagoon & Ceccarelli et al. 2005 \parencite{Ceccarelli2005} & marine & 11    & 0.37  & 11.25 & 0.73 \\
    Pallude Della Rosa Lag Venice & Carreer \& Opitz 1999 \parencite{Carrer1999} & marine & 11    & 0.25  & 0.07  & 0.4 \\
    Patos Lagoon Estuary & Betito 2006 \parencite{Betito2006} & marine & 23    & 0.52  & 17.39 & 0.65 \\
    Peruvian upwelling system 1950s & Jarre-Teichmann 1998 \parencite{Jarre-Teichmann1998} & marine & 19    & 0.43  & 26.81 & 0.7 \\
    Peruvian upwelling system 1960s & Jarre-Teichmann 1998 \parencite{Jarre-Teichmann1998} & marine & 19    & 0.44  & 25.28 & 0.7 \\
    Ria Formosa & Gamito \& Erzini 2005 \parencite{Gamito2005} & freshwater & 13    & 0.65  & 23.39 & 0.58 \\
    Sakumo Lagoon, Ghana & Pauly 2002 \parencite{Pauly2002} & marine & 12    & 0.29  & 35.58 & 0.65 \\
    San Pedro Bay, Leyte, Philippines & Campos 2003 \parencite{Campos2003} & marine & 15    & 0.68  & 9.09  & 0.48 \\
    Seine Estuary & Rybarczyk \& Elkaïm 2003 \parencite{Rybarczyk2003a} & marine & 14    & 0.41  & 20.34 & 0.64 \\
    Sene-Gambia & Samb \& Mendy 2004 \parencite{Samb2003} & marine & 16    & 0.42  & 27.15 & 0.56 \\
    Sierra Leone 1964 & Heymans \& Vakily 2002 \parencite{Heymans2002} & marine & 43    & 0.41  & 6.86  & 0.61 \\
    Sierra Leone 1978 & Heymans \& Vakily 2002 \parencite{Heymans2002}  & marine & 43    & 0.41  & 6.84  & 0.61 \\
    Sierra Leone 1990 & Heymans \& Vakily 2002 \parencite{Heymans2002} & marine & 43    & 0.42  & 6.84  & 0.61 \\
    Sonda Campeche & Manickchand-Heileman et al. 1998 \parencite{ManickchandHeileman1998} & marine & 18    & 0.64  & 6.18  & 0.7 \\
    South Pacific, marine mammals & Morissette & marine & 42    & 0.33  & 7.17  & 0.55  \\
    Southern Brazil & Vasconcellos \& Gasalla 2001 \parencite{Vasconcellos2001} & marine & 12    & 0.49  & 21.66 & 0.73 \\
    Southern Gulf St Lawrence 1980 & Savenkoff et al. 2004 \parencite{Savenkoff2004} & marine & 29    & 0.53  & 2.1   & 0.7 \\
    Southwest Coast Of India & Vivekanandan et al. 2003 \parencite{Vivekanandan2003} & marine & 10    & 0.62  & 16.26 & 0.4 \\
    SriLanka Lake Prakrama Samudra & Moreau et al. 2001 \parencite{Moreau2001} & freshwater & 16    & 0.39  & 71.76 & 0.63 \\
    Strait Of Georgia & Martell et al. 2002 \parencite{Martell2002a} & marine & 26    & 0.47  & 9.81  & 0.73 \\
    Subantartic Plateau New Zealand & Bradford-Grieve et al. 2003 \parencite{Bradford-Grieve2003} & marine & 17    & 0.28  & 64.77 & 0.54 \\
    Tamiahua Lagoon, Golf of Mexico & Abarca-Arenas \& Valero-Pacheco 1993 \parencite{Abarca-Arenas1993} & marine & 12    & 0.63  & 6.46  & 0.7 \\
    Tampa Bay & Walters et al. 2005 \parencite{Walters2005} & marine & 51    & 0.28  & 4.42  & 0.86 \\
    Tampamachoco Lagoon, Mexico & Rosado-Solorzano \& Guzman del Proo 1998 \parencite{Rosado-Solorzano1998} & marine & 22    & 0.36  & 7.95  & 0.7 \\
    Terminos Lagoon, Gulf of Mexico & Manickchand-Heileman et al. 1998 \parencite{ManickchandHeileman1998} & marine & 19    & 0.67  & 3.89  & 0.6 \\
    Terminos Lagoon, seagrass & Rivera-Arriaga et al. \parencite{Rivera-Arriaga2003} & marine & 15    & 0.53  & 10.9  & 0.56 \\
    UK Virgin Islands, Caribbean & Opitz 1996 \parencite{Opitz1996} & marine & 20    & 0.57  & 9.96  & 0.45 \\
    Upper Parana River Floodplain & Angelini \& Agostinho 2005 \parencite{Angelini2005} & freshwater & 39    & 0.21  & 8.95  & 0.82 \\
    Veli Lake, India & Aravindan 1993 \parencite{Aravindan1993} & freshwater & 13    & 0.45  & 32.13 & 0.58 \\
    Weddell Sea & Jarre-Teichmann et al. 1997 \parencite{Jarre-Teichmann1997} & marine & 19    & 0.26  & 4.31  & 0.57 \\
    West Coast of Greenland & Pedersen \& Zeller 2001 \parencite{Pedersen2001} & marine & 21    & 0.52  & 4.98  & 0.74 \\
    West Coast of Sabah & Garces et al. 2003 \parencite{Garces2003} & marine & 28    & 0.49  & 14.18 & 0.6 \\
    West Coast of Sarawak & Garces et al. 2003 \parencite{Garces2003} & marine & 28    & 0.49  & 14.18 & 0.6 \\
    West Coast of Vancouver Island & Martell 2002 \parencite{Martell2002} & marine & 14    & 0.64  & 15.59 & 0.83 \\
    West Greenland, Shrimp Pound & Pedersen 1994 \parencite{Pedersen1994} & marine & 11    & 0.54  & 2.34  & 0.59 \\
    Western Bering Sea & Aydin et al. 2002 \parencite{Aydin2002} & marine & 33    & 0.4   & 3.56  & 0.72 \\
    Western Gulf of Mexico & Arreguin-Sanchez et al. 1993 \parencite{Arreguin-Sanchez1993} & marine & 23    & 0.45  & 5.84  & 0.65 \\
    Yucatan shelf, Gulf of Mexico & Vega-Cendejas \& Arreguin-Sanchez 2001 \parencite{Vega-Cendejas2001} & marine & 20    & 0.53  & 10.33 & 0.63 \\
    \hline
    
\end{longtable}

\subsection{References}
\printbibliography[title=$ $]